\documentstyle[12pt]{article}
\begin{document}

\begin{flushright}
%IMSc-98/11/52,\\
\end{flushright}
   
\vskip 1cm
    
\begin {center}
{\Large \bf Logarithmic correction to the Bekenstein-Hawking entropy of  BTZ black hole}
     
\vskip 1cm
      
{\large  T. R. Govindarajan,  R. K. Kaul 
and V. Suneeta \footnote{e-mail: trg, kaul, suneeta@imsc.ernet.in}}
 
 \vskip 1cm
  
{\small \it $^a$The Institute of Mathematical Sciences,
CIT Campus, Chennai 600113, India}\\

\date{\today}
\end{center}

\vskip 2cm
       
\begin{abstract}
We derive an exact expression for the partition function of  the Euclidean BTZ black hole. Using this, we
show that for a black hole with large horizon area, the correction to the Bekenstein-Hawking 
entropy is  $-3/2~ log(area)$, in agreement with that for the Schwarzschild black 
hole
obtained in the four dimensional canonical gravity formalism and also in 
a Lorentzian computation of BTZ black hole entropy. 
We find that the right expression for the logarithmic correction in the 
context of the BTZ black hole comes from the modular invariance associated
with the toral boundary of the black hole.
\end{abstract}
\newpage
Within the quantum geometry formulation of gravity, it has been argued 
that the quantum degrees of freedom of a (3+1)-dimensional 
black hole can be described in terms of a Chern-Simons
theory on the horizon\cite{ash}, \cite{km1}. For a (3+1)-dimensional
Schwarzschild  black hole, this allows an exact
calculation of the entropy \cite{km1} which for
a large horizon area yields, besides the usual Bekenstein-Hawking
entropy proportional to the area, a next order log(area) correction
with a definite numerical coefficient, $-3/2$ \cite{km2}. $SU(2)$
Wess-Zumino conformal field theory on the boundary  plays an important
role in this calculation. There are other methods which can been
employed to evaluate  black hole entropy.  Exploiting
the nature of corrections to the Cardy formula for density
of states in a two-dimensional conformal field theory, Carlip has 
evaluated  this logarithmic correction in several black hole models
including those from certain string theories\cite{carl2}. Contrary to
the expectations that these corrections may lead to distinguishing
various formulations of quantum gravity, as emphasized by Carlip, 
the coefficient of logarithmic correction to the Bekenstein-Hawking
entropy for large horizon area may have a universal character. 
The same value
of the coefficient appears for a variety of black holes independent
of the dimensions. In particular, the same correction  
was obtained for the entropy of the (2+1)-dimensional Lorentzian BTZ 
black hole \cite{btz} in \cite{carl2} by studying the growth of states in the 
asymptotic conformal field theory at the boundary of the black hole spacetime.
The semi-classical entropy of the BTZ black hole has been earlier obtained in different Lorentzian
\cite{strom}, \cite{carl1} and Euclidean \cite{carlip}, \cite{skg} formulations of gravity.
However, the correction term to semi-classical entropy seen in \cite{carl2} 
has {\em not} been reproduced in the Euclidean path integral calculations for
BTZ black hole, which is surprising. 

In this paper, we derive an {\em exact} expression for the partition function 
of the BTZ black hole in the Euclidean path integral approach. In this
framework, three-dimensional gravity  with a negative cosmological
constant is described in terms of two  $SU(2)$  Chern-Simons theories
 \cite{achtow}, \cite{witten}. Then, 
$SU(2)$  Wess-Zumino conformal field  theories are naturally induced 
on the boundary \cite{witten1}. The quantum degrees of freedom corresponding
to the entropy of the black hole are described by these conformal field theories. 
From the exact expression of the partition function, 
we show that 
there is indeed a correction to the semi-classical entropy that is proportional to the
logarithm of the area (horizon length in this case) with a coefficient 
 $-3/2$ again in agreement with
the result for a four dimensional black hole obtained in ref. \cite{km2}. 
 
The gravity action $I_{grav}$
written in a first-order formalism (using triads $e$ and spin connection 
$\omega$) is the difference of two Chern-Simons actions.
\begin{eqnarray}  
I_{\hbox{\scriptsize grav}}
 ~ = ~I_{\hbox{\scriptsize CS}}[A]~ - ~I_{\hbox{\scriptsize CS}}[\bar A] ,
\label{b9}
\end{eqnarray}
where
\begin{eqnarray}
A ~=~ \left(\omega^a + \frac{i}{l}~ e^a\right) T_a , \qquad
\bar A ~=~ \left(\omega^a - \frac{i}{l}~ e^a\right) T_a
\label{b8}
\end{eqnarray}
are  $\hbox{SL}(2,{\bf C})$ gauge fields (with $T_a=-i\sigma_a/2$).
Here, the negative cosmological constant $\Lambda ~=~ - (1/{l^2})$. 
The Chern-Simons action $I_{\hbox{\scriptsize CS}}[A]$ is 
\begin{eqnarray}
I_{\hbox{\scriptsize CS}} ~=~ {k\over4\pi}\int_M
  \hbox{Tr}\left( A\wedge dA + {2\over3}A\wedge A\wedge A \right)
\label{cs}
\end{eqnarray}
and the Chern-Simons coupling constant is
$k = l/4G$. We note that we can rewrite the Lorentzian gravity action also as a Chern-Simons theory, and
in that case, the Chern-Simons coupling constant $k = - l/4G$; Lorentzian gravity is obtained from the 
Euclidean theory by a continuation $G \rightarrow -G$.

Now, for a manifold with boundary, the Chern-Simons field theory is
described by a Wess-Zumino conformal field theory on the boundary.
Under the decomposition
\begin{eqnarray}
A ~=~ g^{-1}dg ~+~ g^{-1}\tilde A g~ ,
\label{b2}
\end{eqnarray}
the Chern-Simons action (\ref{cs}) becomes \cite{elit}, \cite{ogura}
\begin{eqnarray}
I_{\hbox{\scriptsize CS}}[A]
 ~  = ~I_{\hbox{\scriptsize CS}}[\tilde A]
   ~+~ k I^+_{\hbox{\scriptsize WZW}}[g,\tilde A_z] ~,
\label{b3}
\end{eqnarray}
where $I^+_{\hbox{\scriptsize WZW}}[g,\tilde A_z]$ is the action of
a chiral $SU(2)$ Wess-Zumino model on the boundary $\partial M$,
\begin{eqnarray}
I^+_{\hbox{\scriptsize WZW}}[g,\tilde A_z]
 &=& \frac{1}{4\pi}\int_{\partial M}\hbox{Tr}
 \left(g^{-1}\partial_z g\,g^{-1}\partial_{\bar z} g
 ~-~ 2g^{-1}\partial_{\bar z} g {\tilde A}_z\right) \nonumber \\
 &+& \frac{1}{12\pi}\int_M\hbox{Tr}\left(g^{-1}dg\right)^3 .
\label{b4}
\end{eqnarray}
The `pure gauge' degrees of freedom $g$ are now true
dynamical degrees of freedom at the boundary.

We are interested in computing the entropy of the Euclidean BTZ black hole.
The Euclidean continuation of the BTZ black hole has the topology of a solid torus \cite{cat}.
The metric for the Euclidean BTZ black hole in the usual Schwarzschild-like coordinates is
\begin{eqnarray}
ds^2~=~N^2~d\tau^2 + N^{-2}~dr^2 + r^2~(d\phi + N^{\phi}d\tau)^2
\label{metdef4}
\end{eqnarray}
\noindent where $\tau$ here is the Euclidean time coordinate and 
\begin{eqnarray}
N~=~\left(-M~+~\frac{r^2}{l^2}~-~\frac{J^2}{4r^2}\right)^{\frac{1}{2}}~,
~~~~~~N_{\phi}~=~-\frac{J}{2r^2}
\label{met2}
\end{eqnarray}
The inner and the outer horizons of the Lorentzian black hole solution 
get mapped in the Euclidean continuation to $ir_{-}$ and
$r_{+}$ respectively,  where 
\begin{equation}
r_{\pm}^2~~=~~\frac{Ml^2}{2}~~\left[1~\pm ~\left(1+
\frac{J^2}{M^2~l^2}\right)^{1/2}\right]
\label{hor}
\end{equation}
$M$ and $J$ are the mass and angular momentum of the black hole respectively.

As shown by Carlip and Teitelboim
\cite{cat}, after a coordinate transformation,  
\begin{eqnarray}
x &=& \left({r^2-r_+^2\over r^2-r_-^2}\right)^{1/2}
 \cos\left( {r_+\over l^2}\tau + {|r_-|\over l}\phi \right)
 \exp\left\{ {r_+\over l}\phi - {|r_-|\over l^2}\tau \right\}
 \nonumber \\
 y &=& \left({r^2-r_+^2\over r^2-r_-^2}\right)^{1/2}
\sin\left( {r_+\over l^2}\tau + {|r_-|\over l}\phi \right)
 \exp\left\{ {r_+\over l}\phi - {|r_-|\over l^2}\tau \right\} \\
 z &=& \left({r_+^2-r_-^2\over r^2-r_-^2}\right)^{1/2}
 \exp\left\{ {r_+\over l}\phi - {|r_-|\over l^2}\tau \right\}
 \nonumber
 \label{a11}
 \end{eqnarray}
The black hole metric is just the metric for hyperbolic three-space ${\cal H}_3$
\begin{eqnarray}
ds^2~ =~ \frac{l^2}{z^2} ~(dx^2+dy^2+dz^2),\quad \quad z>0 ,
\label{a4}
\end{eqnarray}
Changing to spherical coordinates
\begin{eqnarray}
x ~=~ R\cos\theta\cos\chi, ~~~~ y~ = ~R\sin\theta\cos\chi, ~~~ z~ =~ R\sin\chi~, 
\label{a6}
\end{eqnarray}
the metric is
\begin{eqnarray}
ds^2 ~=~ \frac{l^2}{ R^2\sin^2\chi}
 ~ \left[dR^2 + R^2d\chi^2 + R^2\cos^2\chi d\theta^2 \right] ,
\label{d1}
\end{eqnarray}
However, one must make global identifications to account for the periodicity of the $\phi$
coordinate in (\ref{metdef4}). These are
\begin{eqnarray}
(\ln R,~\theta,~\chi)
  ~\sim~ (\ln R + \frac{2\pi r_+}{l}, ~\theta + \frac{2\pi|r_-|}{l},~\chi) .
\label{d2}
\end{eqnarray}
The Euclidean BTZ black hole is obtained from hyperbolic space 
${\cal H}_{3}$ by these global identifications.

Using (\ref{b8}), the connection $A^a$  corresponding 
to the metric (\ref{d1}) may be written as: 
\begin{eqnarray}
A^1 ~=~ -\csc\chi ~(d\theta - i\frac{dR}{R}) ,\qquad
A^2~ =~ i\csc\chi ~d\chi ,\qquad
A^3~ =~ i\cot\chi ~(d\theta - i\frac{dR}{R}) .
\label{d3}
\end{eqnarray}
The Chern-Simons formulation of gravity was used to describe the BTZ
black hole first in \cite{mann1}, where
for the Lorentzian black hole, the corresponding gauge fields were
evaluated.

In order to compute the black hole partition function, we first evaluate
the Chern-Simons path integral on a solid torus. This path integral has been
discussed in \cite{elit}, \cite{lab1}, \cite{lab2} and \cite{hayashi}.
Through a suitable  gauge transformation, the connection is set to a 
constant value on the toral boundary. 
In terms of coordinates on the toral boundary $x$ and $y$ with unit period,
we can define $z = (x + \tau y)$ such that 
\begin{eqnarray}
\int_{A} dz~ =~ 1,~~~~~ \int_{B} dz~ =~ \tau
\label{cycle}
\end{eqnarray}
where $A$ is the contractible cycle and $B$ the non-contractible cycle of 
the solid torus and $\tau=\tau_1+i\tau_2$ is the modular parameter of
the boundary torus.
Then, the connection can be written as \cite{lab1}:
\begin{eqnarray}
A ~=~ \left(\frac{-i \pi \tilde u}{ \tau_2}~ d\bar z + \frac{i \pi u}{ \tau_2} 
~dz\right) T_3
\label{adefn}
\end{eqnarray}

\noindent where $u$ and $\tilde u$ are canonically conjugate fields and obey 
the canonical commutation relation:
\begin{eqnarray}
  [\tilde u, u]~ =~ \frac{2 \tau_2}{\pi (k+2)}
\label{ccr}
\end{eqnarray}
They can be related to the black hole parameters by computing the holonomies of 
$A$ around the contractible and non-contractible 
cycles of the solid torus. These holonomies have been computed in \cite{cat} for the general case of a 
rotating BTZ black hole solution with a conical singularity ($\Theta$) at the horizon 
such that the identifications (\ref{d2})
characterizing the black hole now generalize to
\begin{eqnarray}
(\ln R, ~\theta,~ \chi)~ \sim~ (\ln R,~ \theta + ~\Theta, ~\chi) \nonumber \\
(\ln R, ~\theta,~ \chi)  ~\sim~ (\ln R + \frac{2\pi r_+}{l},
~\theta + \frac{2\pi |r_-|}{l}, ~\chi) 
\end{eqnarray}

The former identification corresponds to the $A$ cycle and the latter to the $B$ cycle.
Then the trace of the holonomies around the contractible cycle $A$ and
non-contractible cycle $B$ are:

\begin{equation}
Tr(H_A)~ =~ 2 \cosh (i \Theta), ~~~~~Tr(H_B)~ =~ 2 \cosh \left(\frac{2\pi}{l} (r_+
+ i |r_-|)\right)\label{udefn}
~
\end{equation}

\noindent For the classical black hole solution, $\Theta ~=~ 2\pi$.
From (\ref{adefn}), 
\begin{eqnarray}
A_{z}~ =~ \frac{-i \pi}{\tau_2}~ \tilde u, ~~~~~~~ 
A_{\bar z}~ = ~\frac{i \pi}{\tau_2}~ u
\end{eqnarray}
where
\begin{eqnarray}
u ~=~ \frac{-i}{2 \pi} \left(- i \Theta \tau + \frac{2 \pi 
(r_+ + i |r_-|)}{l}\right),
~~~~~
\tilde u ~=~ \frac{-i}{2 \pi} \left(- i \Theta \bar \tau + \frac{2 \pi 
(r_+ + i |r_-|)}{l}\right)
\end{eqnarray}

\noindent We note here that $\tilde u$ is the canonical conjugate, but not the complex conjugate of $u$. 
This is so  because $A$ is a complex $SU(2)$ connection.

Now, we  write the Chern-Simons path integral on a solid torus with a 
boundary modular parameter $\tau$. For
a fixed boundary value of the connection, i.e. a fixed value of $u$, this path integral is formally
equivalent to a state $\psi_{0}(u, \tau)$ with no Wilson lines in the solid torus. 
The states corresponding  to having closed Wilson lines (along the 
non-contractible cycle) carrying spin $j/2$ 
($j \le k$) representations in the solid torus
are given by \cite{elit}, \cite{lab1}, \cite{lab2},  
\cite{hayashi}:
\begin{eqnarray}
\psi_{j}(u, \tau)~ =~ \exp\left\{ {\pi k\over4\tau_2}\,u^2 \right\}
~\chi_{j}(u, \tau) ,
\label{c3}
\end{eqnarray}
where $\chi_{j} $ are the Weyl-Kac characters for affine $\hbox{SU}(2)$. 
The Weyl-Kac characters can be expressed in terms of the well-known
Theta functions as
\begin{eqnarray}
\chi_{j}(u, \tau)~ =~ \frac{\Theta_{j +1}^{(k+2)}(u, \tau, 0)~-~\Theta_{-j -1}^{(k+2)}(u, \tau, 0)}
{\Theta_{1}^{2}(u, \tau, 0)~-~\Theta_{-1}^{2}(u, \tau, 0)}
\label{c4}
\end{eqnarray}
where Theta functions are given by:
\begin{eqnarray}
\Theta_{\mu}^{k}(u, \tau, z) ~=~ 
\exp (-2 \pi i k z)~ \sum_{n \in \cal Z} \exp 2 \pi i k \left[(n + \frac{\mu}
{2 k})^2 \tau ~+~(n + \frac{\mu}{2 k}) u \right] 
\label{theta}
\end{eqnarray}

The black hole partition function is to be constructed from the boundary state
$\psi_{0}(u, \tau)$. To do that, we
note the following:

a) We must first choose the appropriate ensemble. Here, we choose the microcanonical
ensemble. This corresponds in our picture, to keeping the 
holonomy around the non-contractible
cycle $B$ fixed \cite{brown}. The holonomy around the contractible cycle $A$ is $\Theta$, which has a 
value $2\pi$ for the classical solution. But it is {\em not} held fixed any more, and 
we must sum over contributions to
the partition function from all values of $\Theta$. 
This corresponds to starting with the value of $u$ for the classical solution, 
i.e. with $\Theta = 2\pi$
in (\ref{udefn}), and then considering all other shifts of $u$ of the form
\begin{eqnarray}
u ~\rightarrow~ u + \alpha \tau
\end{eqnarray}
where $\alpha$ is an arbitrary number. This is implemented by a translation operator of the form
\begin{eqnarray}
T ~=~ \exp \left(\alpha \tau \frac{\partial}{\partial u}\right)
\end{eqnarray}
However, this operator is not gauge invariant. The only gauge-invariant way of implementing these
translations is through Verlinde operators of the form
\begin{eqnarray}
W_{j}~ = ~\sum_{n \in \Lambda_{j}} \exp \left(\frac{-n \pi \bar \tau u}{\tau_{2}} + 
\frac{n \tau}{k+2} \frac{\partial}{\partial u} \right)
\end{eqnarray}
where $\Lambda_{j} ~=~ {-j, -j+2,...,j-2, j}$.
This 
means that all possible shifts in $u$ are not allowed, and from considerations of gauge invariance, the
only possible shifts are
\begin{eqnarray}
u ~\rightarrow ~u + \frac{n \tau}{k+2} 
\label{ushift}
\end{eqnarray}
where $n$ is always an integer taking a maximum value of $k$.
Thus, the only allowed values of $\Theta$ are $2\pi(1 + \frac{n}{k+2})$.
We know that acting on the state with no Wilson lines in the solid torus with the
Verlinde operator $W_{j}$
corresponds
to inserting a Wilson line of spin $j/2$ around the non-contractible cycle.
Thus, taking into account all states with different
shifted values of $u$ as in $(\ref{ushift})$ 
means that we have to take into
account all the states  in the boundary corresponding to the insertion of such Wilson lines. These are the
states $\psi_{j}(u, \tau)$ given in (\ref{c3}). 

b) In order to obtain the final partition function, we must integrate over all 
values of the modular parameter, i.e. over all inequivalent tori with the same holonomy around the non-contractible cycle.
The integrand, which is a function of $u$ and $\tau$, must be the square of the partition function of a 
gauged $SU(2)_{k}$ Wess-Zumino model  corresponding to the two $SU(2)$
Chern-Simons theories.  
It must be modular invariant -- modular invariance corresponds to large diffeomorphisms
of the torus, and the partition function must not change under a modular transformation. 

The partition function is then of the form
\begin{eqnarray}
Z ~= ~ \int d\mu(\tau, \bar \tau) ~\left|~\sum_{j=0}^{k} ~a_{j}(\tau)~ \psi_{j}
(u, \tau)~\right|^2
\label{pf}
\end{eqnarray}
where $d\mu(\tau, \bar \tau)$ is the modular invariant measure, and the integration is over a fundamental
domain in the $\tau$ plane. Coefficients $a_{j}(\tau)$ must be chosen  such that the integrand is modular invariant. 
Under an $\mathcal S$ modular transformation, 
$\tau \rightarrow -1/\tau$ and $u \rightarrow u/ \tau$,
the $SU(2)_{k}$ characters transform  as 
\begin{eqnarray}
\chi_{j}(u, \tau) &\rightarrow& \exp \left(-2\pi i k ~\frac{u^2}{4 \tau}\right) 
\chi_{j}\left(\frac{u}{\tau}, ~\frac{-1}{\tau}\right) \nonumber \\
&=& \exp \left(-2 \pi i k ~\frac{u^2}{4 \tau}\right) \sum_{l}~ S_{jl}~\chi_{l}(u, \tau)
\label{chitr}
\end{eqnarray}
where matrix $S_{jl}$ given by
\begin{eqnarray}
S_{jl} ~=~ \sqrt{\frac{2}{k+2}}~ \sin \left[ \frac{\pi (j+1)(l+1)}
{k+2}\right]~ ,~~~~~~~~~ 0 \leq j, ~l \leq k
\end{eqnarray}
We note here that this $S$ matrix is orthogonal: $\sum_{j}~S_{lj}~S_{jp}~=~\delta_{lp}$.

We are interested in the transformation property of the state $\psi_{j}(u, \tau)$ under an $\mathcal S$
modular transformation. The prefactor in (\ref{c3}) transforms into itself under such a transformation
apart from an extra piece that exactly cancels the prefactor in (\ref{chitr}). Thus, under an $\mathcal S$
transformation ($\tau \rightarrow -1/\tau$),
\begin{eqnarray}
\psi_{j}(u, \tau) ~\rightarrow~ \sum_{l}S_{jl} ~\psi_{l}(u, \tau)
\end{eqnarray}
Under a $\mathcal T$ modular transformation ($\tau \rightarrow \tau
+1$), $\psi_{j}(u, \tau)$ picks up a phase,
\begin{eqnarray}
\psi_{j}(u, \tau) ~\rightarrow~ \exp (2\pi i m_{j})~ \psi_{j}(u, \tau)
\end{eqnarray}
where $m_{j}~=~\frac{(j+1)^2}{2(k+2)}~-~ \frac{1}{4}$.
For the integrand in (\ref{pf}) to be modular invariant, the coefficient $a_{j}(\tau)$ must
transform under the $\mathcal S$ transformation as
$a_{j}(\tau) ~\rightarrow~ \sum_{p}a_{p}(\tau)~ S_{pj}$
and under the $T$ transformation as 
$a_{j}(\tau)~ \rightarrow ~\exp (-2\pi i m_{j}) a_{j}~(\tau)$.
Further, since the integrand must correspond to the square of the partition function of a gauged 
$SU(2)_{k}$ Wess-Zumino model, the coefficients $a_{j}(\tau)$ are just the 
complex conjugate of $SU(2)_{k}$
characters corresponding to $u~=~0$, i.e $(\psi_{j}(0, \tau))^{*}$. 
The black hole partition function therefore is
\begin{eqnarray}
Z_{bh} = \int d\mu(\tau, \bar \tau) \left|~\sum_{j=0}^{k} 
~(\psi_{j}(0, \tau))^{*} ~\psi_{j}(u, \tau)~\right|^2
\label{bhpf}
\end{eqnarray}
Finally the modular invariant measure is  
\begin{eqnarray}
d\mu(\tau, \bar \tau)~ =~ \frac{d\tau d\bar \tau}{\tau_{2}^{2}}
\end{eqnarray}

The expression (\ref{bhpf}) is an {\em exact} expression for the partition 
function of Euclidean
black hole. To make a comparison with  the semi-classical entropy of  
black hole, 
we evaluate the expression (\ref{bhpf}) for large  horizon radius $r_{+}$
by the saddle-point method.
Substituting from (\ref{c3}), (\ref{c4}) and (\ref{theta}),
the saddle point of the integrand occurs when $\tau_{2}$ is 
proportional to $r_{+}$ and therefore large. But for $\tau_{2}$ large, the character $\chi_{j}$ is
\begin{eqnarray}
\chi_{j}(\tau, u)~ \sim~ \exp \left[\frac{\pi i \left(\frac{(j+1)^2}{k+2} - 
\frac{1}{2}\right)}{2} \tau \right] ~\frac{\sin \pi (j+1)u}{\sin \pi u}
\label{asym}
\end{eqnarray}
We now use in (\ref{bhpf}) the form of the character for large $\tau_{2}$  from (\ref{asym}).
In the expression for $u$ in (\ref{udefn}), we replace
$\Theta$ by its classical value $2\pi$. The computation has been done
with positive coupling constant $k$ and at the end, 
we must perform an analytic continuation to the Lorentzian black hole, by taking
$G \rightarrow -G$. It can be checked that after the analytic continuation, it is the 
spin $j=0$ in the sum over characters in (\ref{bhpf}) that dominates the partition function.
 
We obtain the leading behaviour of the partition function (\ref{bhpf}) for large $r_{+}$ (and
when $|r_{-}|<< r_{+}$) by first performing the integration over $\tau_{1}$ in this regime. 
The $\tau_{2}$ integration is done by the method of steepest descent. The saddle-point is at
$\tau_{2}~=~r_{+}/l$.
Expanding around the saddle-point,
by writing $\tau_{2}~=~r_{+}/l ~+~ x $ and then  integrating over $x$, we obtain 
\begin{eqnarray}
Z_{bh} ~=~\frac{l^2}{r_{+}^2} ~\exp \left(\frac{- 2\pi k r_{+}}{l}\right) ~\int dx 
~\exp \left[-\frac{\pi k l}{2 r_{+}}~ x^2 \right]
\end{eqnarray}
The integration produces a factor proportional to $\sqrt{r_+}$. 
The 
partition function for the Lorentzian black hole
of large horizon area $2 \pi r_{+}$ after the analytic continuation $G \rightarrow -G$ is then
\begin{eqnarray}
Z_{Lbh} ~=~ \frac{l^2}{r_{+}^2}~ \sqrt{\frac{8 r_{+} G}{\pi l^2}} ~\exp 
\left(\frac{2\pi r_{+}}{4 G}\right) 
\label{spbhpf}
\end{eqnarray}
upto a multiplicative constant. The logarithm of this expression 
yields the black hole entropy for large horizon length $r_{+}$:
\begin{eqnarray}
S~ = ~\frac{2 \pi r_{+}}{4 G}~ - ~\frac{3}{2} \log \left({\frac{2 \pi r_{+}}
{4G}}\right)~ +~ .~.~.~.
\label{entropy}
\end{eqnarray}

The leading
contribution to the black hole entropy
is the familiar Bekenstein-Hawking term.
The next-order correction to the semi-classical entropy is the logarithm of 
the black hole area $2\pi r_{+}$.
The coefficient $- 3/2$  of this correction is in agreement with that
of the logarithmic correction of  
semi-classical entropy of  
four dimensional Schwarzschild black hole first observed in ref. \cite{km2} 
in the quantum geometry formulation
of gravity. 
The semi-classical Bekenstein-Hawking entropy for Euclidean BTZ black hole was
previously studied  in the path integral formulation in ref. \cite{carlip}, 
but the logarithmic correction was not seen there. As described above, the
right logarithmic correction is obtained  by considering the correct 
modular invariant measure while integrating over all inequivalent tori
(as the holonomy around the non-contractible cycle is held fixed).

The calculation presented here should be contrasted with an earlier calculation of partition function
of a BTZ black hole coupled to a scalar field \cite{mann2}. This is a perturbative one-loop 
calculation which incorporates a specific type of fluctuation, namely a scalar field.
For small $r_{+}$, this leads to a different coefficient of the 
the logarithmic correction in the entropy. 
On the other hand, our calculation is exact; it includes all possible quantum gravity fluctuations. It is 
therefore not surprising that the results differ.

Finally, we make a few remarks on the $AdS$ gas partition function. 
As is well known \cite{ms},
the action for the $AdS$ gas can be obtained from that of the BTZ black hole by a transformation.
For the case of a non-rotating black hole, this transformation has the form $r_{+}/l \rightarrow
l/r_{+}$.
With this change, the $AdS$ gas partition function is
\begin{eqnarray}
Z_{AdS}[r_{+}] = \int d\mu(\tau, \bar \tau) \left|~\sum_{j=0}^{k}
~(\psi_{j}(0, \tau))^{*} ~\psi_{j}(u', \tau)~\right|^2
\label{adspf}
\end{eqnarray}
where 
$u' ~=~ \frac{-i}{2 \pi} \left(- i 2\pi \tau + \frac{2 \pi
l}{r_{+}}\right)$.

The $AdS$ gas partition function can again be evaluated by saddle-point method. Small $r_{+}$
leads to a saddle-point with $\tau_{2}$ large. In this limit of small $r_{+}$ (i.e small temperature),
the partition function is
\begin{eqnarray}
Z_{AdS}[r_{+}] = (\frac{r_{+}}{l})^\frac{3}{2}  ~\exp
\left(\frac{ 2\pi l^2 }{4 r_{+} G}\right) 
\end{eqnarray}
This, at the leading order, agrees with the corresponding expression obtained in ref. \cite{ms}.

\vskip 1cm
V.S. would like to  thank S. Carlip for useful discussions. We would also like to thank N. D. Hari Dass
for comments and discussions on the results of our paper.


\begin{thebibliography}{100}

\bibitem{ash} A. Ashtekar, J. Baez, A. Corichi and K. Krasnov,  
	      Phys.Rev.Lett.{\bf 80}, 904 (1998).
\bibitem{km1} R. K. Kaul and P. Majumdar, Phys. Lett. {\bf B439}, 267 (1998).
\bibitem{km2} R. K. Kaul and P. Majumdar, Phys. Rev. Lett. {\bf 84}, 5255 
(2000); \\
S. Das, R. K. Kaul and P. Majumdar, Phys. Rev. {\bf D63}, 044019 (2001).
\bibitem{btz}  M. Banados, C. Teitelboim and J. Zanelli,
                Phys.Rev.Lett. {\bf 69}, 1849 (1992); \\
M. Banados, M. Henneaux, C. Teitelboim and J. Zanelli, Phys. Rev. {\bf D48}, 1506 (1993).
\bibitem{carl2} S. Carlip, Class. Quant. Grav. {\bf 17}, 4175 (2000).
\bibitem{strom}  A. Strominger, JHEP {\bf 9802}, 009 (1998).
\bibitem{carl1} S. Carlip, Nucl. Phys. Proc. Suppl. {\bf 88}, 10 (2000).
\bibitem{carlip} S. Carlip, Phys. Rev. {\bf D55}, 878 (1997).
\bibitem{skg} V. Suneeta, R. K. Kaul and T. R. Govindarajan, Mod.Phys.Lett.A 
{\bf 14}, 349 (1999).
\bibitem{achtow} A. Achucarro and P. Townsend, Phys.Lett. B {\bf 180}, 89 (1986).
\bibitem{witten} E. Witten, Nucl.Phys.B {\bf 311}, 46 (1988);\\
S. Carlip, Class. Quant. Grav. {\bf 16}, 3327 (1999).
\bibitem{witten1} E. Witten, Commun. Math. Phys. {\bf 121}, 351 (1989);
\\
R.K. Kaul, Commun. Math. Phys. {\bf 162}, 289 (1994);\\
R.K. Kaul, Chern-Simons theory, knot invariants, vertex models and three
manifold invariants, hep-th/9804122,  published in {\it
Frontiers of Field Theory, Quantum Gravity and Strings (Horizons in
World Physics, Vol. 227)}, NOVA Science Publishers, New York, (1999).
\bibitem{mann1} D. Cangemi, M. Leblanc and R. Mann, Phys. Rev. {\bf D48}, 3606 (1993).
\bibitem{elit} S. Elitzur, G. Moore, A. Schwimmer and N. Seiberg, Nucl. Phys. B {\bf 326}, 108 (1989).
\bibitem{ogura} W. Ogura, Phys. Lett. {\bf B229}, 61 (1989);\\
S. Carlip, Nucl. Phys. B {\bf 362}, 111 (1991).
\bibitem{cat} S. Carlip and C. Teitelboim, Phys. Rev. {\bf D51}, 622 (1995).
\bibitem{lab1} J. M. F. Labastida and A. V. Ramallo, Phys. Lett. {\bf B 227}, 
92 (1989).
\bibitem{lab2} J. M. Isidro, J. M. F. Labastida and A. V. Ramallo, Nucl. Phys. B {\bf 398}, 187 (1993).
\bibitem{hayashi} N. Hayashi, Prog. Theor. Phys. Suppl. {\bf 114}, 125 (1993).
\bibitem{brown} J. D. Brown, G. L. Comer, E. A. Martinez, J. Melmed, B. F. Whiting and J. W. York,
Class. Quant. Grav. {\bf 7}, 1433 (1990).
\bibitem{mann2} R. Mann and S. Solodukhin, Phys. Rev. {\bf D55}, 3622 (1997).
\bibitem{ms} J. Maldacena and A. Strominger, JHEP {\bf 9812} 005 (1998).
\end{thebibliography}
\end{document}